\documentclass[pdflatex,sn-aps]{sn-jnl}

\usepackage{graphicx}%
\usepackage{multirow}%
\usepackage{amsmath,amssymb,amsfonts}%
\usepackage{amsthm}%
\usepackage{mathrsfs}%
\usepackage[title]{appendix}%
\usepackage{xcolor}%
\usepackage{textcomp}%
\usepackage{manyfoot}%
\usepackage{booktabs}%
\usepackage{algorithm}%
\usepackage{algorithmicx}%
\usepackage{algpseudocode}%
\usepackage{listings}
\usepackage{tabto}

\begin{document}
\begin{flushright}
DOI: 10.17181/hvcpv-bk752
\end{flushright}
\title[Article Title]{A vertex fitting package}

\vspace*{0.5cm}
\author*[1]{\sur{Franco Bedeschi}}\email{franco.bedeschi@pi.infn.it}

\affil*[1]{\orgname{INFN - Sezione di Pisa}, \orgaddress{\street{Largo B. Pontecorvo 3}, \city{Pisa}, \postcode{56127}, \country{Italy}}}

\maketitle

\section{Introduction}\label{sec1}

Vertex fitting code is commonly found within the analysis packages of several HEP experiments, unfortunately it usually deeply packaged inside their software infrastructure, making it cumbersome to use in the context of external applications such, for instance, in the study of the performance of the proposed FCCee detectors. In this note a totally independent  package is described. The only dependencies being the ROOT libraries~\cite{Brun:1997pa, Antcheva:2011zz}, making it easy to use in a wide range of applications. This code has already been used in several studies of physics at the future FCC electron-positron accelerator; see for instance~\cite{Aleksan:2024jwm}.

The code currently works with track trajectories that are either helices or straight lines, as generated by charged or neutral particles in a constant magnetic field, but is expandable in principle  to other type of trajectories or parameterizations.

This vertexing package,  in addition to fitting a vertex given a list of tracks, supports several features: addition and subtraction of tracks, momenta and their errors of the tracks at the vertex point, external vertex constraints, mass constraints, and the capability to perform fits of chain decays. 

This note is organized as follows: in section~\ref{sec2} the basic formulas for the fit are derived, section~\ref{sec3} describes how to use the software and two examples are given in section~\ref{sec4}. Details on the fit formulas derivation are shown in appendix~\ref{secA}. Several formulas needed for trajectories in a uniform magnetic field, track parameters and related derivatives are contained in appendix~\ref{secB} and~\ref{secC}.

\section{Basic formulas}\label{sec2}
Two main approaches are used for vertex fitting: the simplest finds the location of the point closest to all tracks without changing the track parameters; the second is more complex and forces all tracks to go through the same point by changing the track parameters. We look now at the first method, while the second will be developed in the next subsection.
 
\subsection{Vertex fit without track parameter steering}\label{subsec2.1}
Given a track trajectory $\vec{x}(s;\vec{\alpha})$,  where $\vec{\alpha}$ are the track parameters and the phase, $s$, defines the position on the  trajectory, the covariance matrix $S$ of a point on the track at fixed $s$ is given by 
\[ S = W^{-1} = \frac{\partial\vec{x}}{\partial\vec{\alpha}}C\left(\frac{\partial\vec{x}}{\partial\vec{\alpha}}\right)^t = A C A^t \]
where $C$ is the covariance matrix of the track parameters $\vec{\alpha}$. The $\chi^2$ to minimize is the following:
\begin{equation}
\label{Chi2NoSteer}
\chi^2 = \sum_{i=1}^N (\vec{x}(s_i;\vec{\alpha}_i)-\vec{x}_V)^t W_i(\vec{x}(s_i;\vec{\alpha}_i)-\vec{x}_V)
\end{equation}
where $\vec{x}_V$ is the vertex to be determined. The unknown parameters in this fit, besides the vertex position, $\vec{x}_V$, are the phases, $s_i$. We now set to 0 the derivatives of the $\chi^2$, after using a first order approximation for $\vec{x}(s; \vec{\alpha}) \simeq \vec{x}^0+\frac{\partial\vec{x}}{\partial s}\delta s = \vec{x}^0+\vec{a}\delta s$:
\begin{equation}
\frac{1}{2}\frac{\partial\chi^2}{\partial s_k} = \vec{a}_k^t W_k(\vec{x}^0_k+\vec{a}_k\,\delta s_k-\vec{x}_V)=
\vec{a}_k^tW_k(\vec{x}^0_k-\vec{x}_V)+\vec{a}_k^tW_k\vec{a}_k\,\delta s_k = 0
\end{equation}
which can be solved for $\delta s_k$:
\begin{equation}
\label{dsNoSteer}
\delta s_k = \frac{\vec{a}_k^tW_k(\vec{x}_V-\vec{x}^0_k)}{\vec{a}_k^tW_k\vec{a}_k} =
\frac{\vec{a}_k^t}{a_k}W_k(\vec{x}_V-\vec{x}^0_k)
\end{equation}
then:
\begin{equation}
\frac{1}{2}\frac{\partial\chi^2}{\partial \vec{x}_V} =
-\sum_{i=1}^N W_i(\vec{x}^0_i+\vec{a}_i\,\delta s_i -\vec{x}_V) =
-\sum_{i=1}^N W_i(\vec{x}^0_i+\frac{\vec{a}_i\vec{a}_k^t}{a_k}W_i(\vec{x}_V-\vec{x}^0_i)
-\vec{x}_V)
\end{equation}
that can be solved for $\vec{x}_V$:
\begin{equation}
\label{XvNoSteer}
\vec{x}_V = (\sum_{i=1}^N D_i)^{-1}(\sum_{i=1}^N D_i \vec{x}^0_i) 
= D^{-1}(\sum_{i=1}^N D_i \vec{x}^0_i)
\end{equation}
where:
\begin{equation}
D_i = W_i-W_i\frac{\vec{a}_i\vec{a}_i^t}{a_i}W_i
\end{equation}
The error matrix on $\vec{x}_V$ is obtained by error propagation on the $\vec{x}^0_i$:
\begin{equation}
Cov(\vec{x}_V) = D^{-1}(\sum_{i=1}^N D_iW^{-1}_iD_i)D^{-1}
\end{equation}

The procedure can be iterated by using $\vec{x}_V$ as obtained in eq.~\ref{XvNoSteer} to update the $s_i$ (from eq.~\ref{dsNoSteer}) and then the $\vec{x}_i^{\,0}$.

\subsection{Vertex fit with parameter steering}{\label{subsec2.2}
	The method described in section~\ref{subsec2.1} is very fast and provides a reliable vertex position and error matrix, however it is not suitable to  allow further functionality, such as mass constraints or applications in chain decays. We then describe another method where $N$ input tracks are forced to pass through a common vertex by varying their track parameters.
Let 
\[ \vec{x}_i(s_i,\vec{\alpha}_i)\simeq \vec{x}_i(s'_i, \vec{\alpha}'_i)+\frac{\partial \vec{x}_i}{\partial \vec{\alpha}_i}\,(\vec{\alpha}_i-\vec{\alpha}'_i)+\frac{\partial \vec{x}_i}{\partial\,s_i}\,(s_i-s'_i) = \vec{x}_i^{\,0}+A_i\delta\vec{\alpha}_i+\vec{a}_i\delta\,s_i \]

be all track equations and their first order expansion in the track parameters, $\vec{\alpha}_i$  and the variables describing the position in each trajectory, $s_i$. The symbol $'$ is used to indicate the expansion point. Let also $C_i$ be the covariance matrices of the original track parameters $\vec{\alpha}_i^{\,0}$
	The vertex fit amounts to minimizing the following $\chi^2$:

\begin{equation}
\label{Chi2Steer}
\chi^2=\sum_{i=1}^N (\vec{\alpha}_i-\vec{\alpha}^{\,0}_i)^{\,t}C_i^{-1}(\vec{\alpha}_i-\vec{\alpha}^{\,0}_i) + 2(\vec{x}(s_i, \vec{\alpha}_i)-\vec{x}_V)^t\vec{\lambda}_i\}
\end{equation}
where $\vec{\alpha}_i$ and $s_i$ are the modified track parameters and the phases at the vertex, $\vec{x}_V$ is the vertex to be determined and $\vec{\lambda}_i$ are Lagrange multipliers. The fit varies the track parameters and the phases until  all tracks go through a common vertex $\vec{x}_V$.  
	The solution is found to be:

\begin{equation}
\vec{x}_V = \left(\sum_{i=1}^N  D_i\right)^{-1}
\left(\sum_{i=1}^N  D_i(\vec{x}_i^{\,0}+A_i\,\delta\vec{\alpha}^{\,0}_i)\right) = D^{-1}\left(\sum_{i=1}^N  D_i(\vec{x}_i^{\,0}+A_i\,\delta\vec{\alpha}^{\,0}_i)\right)
\end{equation}
where $\delta\vec{\alpha}_i^{\,0} = \vec{\alpha}_i^{\,0}-\vec{\alpha}_i$ is the difference between the input and final track parameters.
The corresponding error matrix on $\vec{x}_V$ is obtained by taking the average of the fluctuations of $\vec{x}_V$ due to those of the track parameters, $\vec{\alpha}^{\,0}_i$:
\begin{equation}
\begin{array}{lcl}
C_V = Cov(\vec{x}_V) &=& 
D^{-1}
\left(\sum_{ij}  D_iA_i
<\delta\vec{\alpha}_i^{\,0}\delta\vec{\alpha}_j^{\,0t}>A_i^{\,t}
D_j\right)
D^{-1}\\
&=&
D^{-1}
\left(\sum_{i}  D_i W_i^{-1} D_i\right)
D^{-1}
\end{array}
\end{equation}

We note that the vertex position and its errors are almost the same as obtained with the simpler fit with no parameter steering, however here we have the updated parameters satisfying the  vertex constraint. This is very useful to expand the functionality of the vertex fit as seen in later sections, when dealing with complex constraints or chain decays.

\subsection{$\chi^2$ and updated track parameter errors}
Since the constraints are all zero after solving the minimization problem, the $\chi^2$ is given only by the sum of the variation of the track parameters:
\begin{equation}
\chi^2 = \sum_{i=1}^N (\vec{\alpha}_i-\vec{\alpha}^{\,0}_i)^{\,t}C^{-1}_i (\vec{\alpha}_i -\vec{\alpha}^{\,0}_i)= 
\sum_{i=1}^N \vec{\lambda}_i^{\,t}(A_iC_iA_i^t)\vec{\lambda}_i =
\sum_{i=1}^N \vec{\lambda}_i^{\,t}W_i^{-1}\vec{\lambda}_i
\end{equation}

Comparing with eq.~\ref{Chi2NoSteer} the Lagrange multipliers correspond to the vector distances of the tracks from the vertex. Given the structure of the $\chi^2$ it is easy to separate the contribution of each track. The $\vec{\lambda}_i$ can be obtained directly from eq.~\ref{lambda}:

\begin{equation}
\vec{\lambda}_i = D_i(\vec{x}^0_i+A_i\,\delta\vec{\alpha}^{\,0}_i-\vec{x}_V) = 
D_i\sum_{k=1}^N (I\delta_{ik}-D^{-1}D_k)(\vec{x}^0_k+A_i\,\delta\vec{\alpha}^{\,0}_i)
\end{equation}

This leads to a closed form for the  relation between the updated track parameters, $\vec{\alpha}_i$,  and those in input, $\vec{\alpha}_i^0$:

\begin{eqnarray}
\nonumber
\vec{\alpha}_i &=&\vec{\alpha}_i^0-C_i\,A_i^{\,t}\vec{\lambda}_i\\
&=&\vec{\alpha}_i^0-C_i\,A_i^{\,t}
D_i\sum_{k=1}^N (I\delta_{ik}-D^{-1}D_k)(\vec{x}^0_k+A_k\,\delta\vec{\alpha}^{\,0}_k)
\end{eqnarray}
so:
\begin{eqnarray}
\nonumber
d\vec{\alpha}_i &=&d\vec{\alpha}_i^0-C_i\,A_i^{\,t}
D_i\sum_{k=1}^N (I\delta_{ik}-D^{-1}D_k)A_k\,d\vec{\alpha}^{\,0}_k\\
&=&\sum_{k=1}^N [I_5\delta_{ik} -C_i\,A_i^{\,t}
D_i(I_3\delta_{ik}-D^{-1}D_k)A_k]\,d\vec{\alpha}^{\,0}_k\\
&=&\sum_{k=1}^N M^i_k\,d\vec{\alpha}^{\,0}_k
\end{eqnarray}
where the subscript under the $I$ indicates the dimensionality of the unit matrix. So 
\begin{equation}
\label{Mik}
M^i_k = \frac{\partial \vec{\alpha}_i}{\partial\vec{\alpha}^0_k}
\end{equation}
Finally the errors on the updated  track parameters are given by:

\begin{equation}
\label{covAij}
<\delta\vec{\alpha}_i\delta\vec{\alpha}_j^{\,t}> =
\sum_{k=1}^N M^i_k<\delta\vec{\alpha}_k^0\delta\vec{\alpha}_k^{0\,t}>
M^{j\,t}_k = \sum_{k=1}^N M^i_k C_k (M^j_k)^{\,t}
\end{equation}

\subsection{Momenta at vertex and their errors}
The track momentum at the vertex can be calculated from the updated track parameters and the phase: 
$\vec{p}_i = \vec{p}_i(\vec{\alpha}_i, s_i)$ therefore:
\begin{equation}
\nonumber
\frac{\partial\vec{p}_i}  {\partial\vec{\alpha}_k^0} = 
\frac{\partial\vec{p}_i}  {\partial\vec{\alpha}_i}
\frac{\partial\vec{\alpha}_i}  {\partial\vec{\alpha}^0_k} + 
\frac{\partial \vec{p}_i}  {\partial s_i}
\frac{\partial s_i}  {\partial \vec{\alpha}^0_k}
\end{equation}
where the phase derivatives are given by:
\begin{equation}
\frac{\partial s_i}{d\vec{\alpha}^{\,0}_k} = \vec{S^i_k}^{\,t} =
\frac{\vec{a}_i^{\,t}W_i}{a_i}(D^{-1}D_k-I\delta_{ik})A_k
\end{equation}

Using a similar strategy as that used to calculate $<\delta\vec{\alpha}_i\delta\vec{\alpha}_j^{\,t}>$, we get all the $<\delta\vec{p}_i\delta\vec{p}_j^t>$ terms:
\begin{equation}
C_{ij} = <\delta\vec{p}_i\delta\vec{p}_j^{\,t}> = 
\sum_{k=1}^N \frac{\partial\vec{p}_i}  {\partial\vec{\alpha}_k^0} C_k
\left(\frac{\partial\vec{p}_i}  {\partial\vec{\alpha}_k^0}\right)^t\mbox{\;\;i, j = 1, N}
\end{equation}

This can be used to get the covariance matrix of the total vertex momentum 
$\vec{p} = \sum_i \vec{p}_i$:
\begin{equation}
<\delta\vec{p}\delta\vec{p}^{\,t}> = \sum_{i,j}^N C_{ij}
\end{equation}

The correlation between the momenta and the vertex position is given by:
\begin{equation}
C_{i,0} = <\delta\vec{p}_i\delta{\vec{x}^{\,t}}> = 
\sum_{k=1}^N \frac{\partial\vec{p}_i}  {\partial\vec{\alpha}_k^0} C_k \left(X_k\right)^t
\end{equation}
where $X_k = D^{-1}A_k$.

\subsection{Covariance matrix of vertex track parameters}
This code has the feature of using each found vertex as an additional track (charged or neutral) that can be used for vertexing. This requires to calculate the vertex track parameters, that is trivial given the position of the vertex and its total momentum, $\vec{\alpha}_V(\vec{x}, \vec{p})$, and their covariance matrix, $C_V$.  

It is useful to define a global vector:
\begin{equation}
\vec{q} = \left(
\begin{array}{c}
\vec{x}\\
\vec{p}_1\\
\cdots\\
\vec{p}_N
\end{array}\right)
\end{equation}
and its covariance matrix:
\begin{equation}
Q = Cov(\vec{q}) =\left(
\begin{array}{llcl}
C_{00}&C_{01}&\cdots&C_{0N}\\
C_{10}&C_{11}&\cdots&C_{1N}\\
\cdots&\cdots&\cdots&\cdots\\
C_{N0}&C_{N1}&\cdots&C_{NN}
\end{array}\right)
\end{equation}
Then, after noting that $\partial \vec{\alpha}_V/\partial\vec{p}_i = \partial \vec{\alpha}_V/\partial\vec{p}$:
\begin{equation}
C_V = \frac{\partial \vec{\alpha}_V}{\partial \vec{q}} Q
\left(\frac{\partial \vec{\alpha}_V}{\partial \vec{q}}\right)^t
\end{equation}

\subsection{External constraint}\label{subsec2.3}
Adding an external constraint to the $\chi^2$ in the form 
$(\vec{x}_V-\vec{y})^tV^{-1}(\vec{x}_V-\vec{y})$, where $\vec{y}$ is an independent measurement of the vertex and $V$ its covariance matrix, is relatively straightforward. Indeed all derivatives are the same except for those with respect to $\vec{x}_V$ that become:
\[ \frac{1}{2}\frac{\partial\chi^2}{\partial\,\vec{x}_V}  =
-\sum_{i=1}^N\vec{\lambda}_i + V^{-1}(\vec{x}_V-\vec{y}) 
= \sum_{i=1}^N  D_i(\vec{x}_V-\vec{x}_i^{\,0}-A_i\,\delta\vec{\alpha}^{\,0}_i)
+ V^{-1}(\vec{x}_V-\vec{y})= 0 \]
solving this the final vertex $\vec{x}_V$ is:
\begin{equation}
\vec{x}_V = \left(\sum_{i=1}^N  D_i+V^{-1}\right)^{-1}
\left(V^{-1}\vec{y}+\sum_{i=1}^N  D_i(\vec{x}_i^{\,0}+A_i\,\delta\vec{\alpha}^{\,0}_i)
\right) = D_V^{-1}\left(V^{-1}\vec{y}+
\sum_{i=1}^N  D_i(\vec{x}_i^{\,0}+A_i\,\delta\vec{\alpha}^{\,0}_i)\right)
\end{equation}
Its covariance matrix can be easily  obtained by error propagation:
\begin{equation}
C_V = Cov(\vec{x}_V)=
D_V^{-1}\left(V^{-1}+\sum_{i=1}^N D_iW_i^{-1}D_i\right)D_V^{-1}
\end{equation}

The effect of this type of contraint on all above formulas is just a modification of the matrix $D$:
\begin{equation}
D = \sum_i D_i \rightarrow D =  \sum_i D_i+V^{-1}
\end{equation}

\subsection{Generic constraint}
Both vertex and moments can be improved applying a constraint, for instance an invariant mass constraint on a subset of particles used for vertexing. We solve here the case of a generic set of contraints using the notations defined in the previous sections. 

Let $\vec{f}(\vec{q})$ be  the constraints, then the $\chi^2$ is given by:
\begin{equation}
\chi^2 = (\vec{q}-\vec{q}_0)^t Q^{-1}(\vec{q}-\vec{q}_0)+2\vec{f}^{\,t}(\vec{q})\vec{\lambda}
\end{equation}
We linearize the constraint around a point $\vec{q}\,'$ to allow the solution by iteration,
\begin{equation}
\vec{f}(\vec{q}) \simeq \vec{f}(\vec{q}\,')+\frac{\partial \vec{f}}{\partial\vec{q}}(\vec{q}-\vec{q}\,')
= \vec{f}_0 + B\delta\vec{q}
\end{equation}
we also define $\delta\vec{q}_0 = \vec{q}_0-\vec{q}\,'$. The linearized problem is solved by setting to zero the derivatives of the $\chi^2$  with respect to the vectors $\vec{p}$ and $\vec{\lambda}$:
\begin{equation}
\frac{1}{2}\frac{\partial\chi^2}{\partial\vec{q}} =
Q^{-1}(\delta\vec{q}-\delta{q}_0)+B^t\lambda = 0
\;\;\rightarrow\;\;\delta\vec{q} = \delta{q}_0-QB^t\lambda
\end{equation}
Replacing $\delta\vec{q}$ in the constraint we obtain $\lambda$:
\begin{equation}
\vec{f}_0+B\delta\vec{q}_0-BQB^t\lambda = 0
\;\;\rightarrow\;\;\lambda = (BQB^t)^{-1}(\vec{f}_0+B\delta\vec{q}_0)
\end{equation}
and finally :
\begin{equation}
\delta\vec{q} = \delta\vec{q}_0-QB^t(BQB^t)^{-1}(\vec{f}_0+B\delta\vec{q}_0)
 = (I-QB^t(BQB^t)^{-1}B)\delta{q}_0-QB^t(BQB^t)^{-1}\vec{f}_0
\end{equation}
The covariance of the updated parameters is obtained averaging over the fluctuations of the $\vec{q}_0$:
\begin{equation}
Cov(\vec{q}) =  (I-QB^t(BQB^t)^{-1}B)Q (I-QB^t(BQB^t)^{-1}B)^t = Q-QB^t(BQB^t)^{-1}BQ
\end{equation} 
\newpage
\section{Software implementation}\label{sec3}
The vertex fitting code is implemented with three classes: \texttt{VertexFit}, that performs the vertex fit with some options, \texttt{VertexMore}, that provides additional features such as update the fit with mass constraints, calculation of track momenta at the vertex and their errors and calculation of the vertex track parameters to allow use in chain decays, while some usuful general formulas are kept in \texttt{TrkUtil}.These three classes are currently part of the \texttt{DELPHES} package~\cite{deFavereau:2013fsa, Selvaggi:2014mya, Selvaggi:2016ydq} available at \texttt{https://github.com/delphes/delphes}. These three classes are self-contained and the only other dependencies are with the standard \texttt{ROOT} package \cite{Brun:1997pa, Antcheva:2011zz}.

	It is worth noting that, while \texttt{VertexFit} can be used with any units or magnetic field, since its formulas depend only on the track parameters, this is not the case for \texttt{VertexMore} since explicit track parameter to momenta conversions are needed. Meters, GeV and B-field of 2 Tesla are the default. An option to switch to millimeters is provided in the constructor, but at present the field needs to be changed by hand in the code.

Given an array of pointers to track parameter sets, \texttt{tPar}, and their covariance matrices, \texttt{tCov}, the vertex fit is obtained by the following simple code sequence:\\

\noindent\texttt{const Int\_t N = something;\\
TVectorD* tPar[N]; TMatrixDSym* tCov[N];\\
.... assign tPar[i] and tCov[i] .... \\
eg: tPar[i] = new TVectorD(par); tCov[i] = new TMatrixDSym(cov);\\
...........................................\\
VertexFit* Vfit = new VertexFit(N, tPar, tCov);\\
}
This default constructor assumes that all input tracks are charged. If this is not the case, one needs to tell the fit which tracks are charged and which are not by adding a \texttt{Bool\_t} array that is \texttt{kTRUE} for charged tracks and \texttt{kFALSE} for neutrals as follows:\\

\noindent
\texttt{Bool\_t Charge[N];\\
... assign Charge[i] values ...\\
VertexFit* Vfit = new VertexFit(N, tPar, tCov, Charge);}\\

The fit results are provided by the following methods of \texttt{VertexFit}:\\

\noindent
\texttt{TVectorD XvFit = Vfit->GetVtx();}\tab// Vertex position\\
\texttt{TMatrixDSym XvCov = Vfit->GetVtxCov();}// Vertex position covariance\\
\texttt{Int\_t Ntr = Vfit->GetNtr();}\tab// Number of tracks in fit\\
\texttt{Double\_t Chi2 = Vfit->GetVtxChi2();}\tab// Vertex $\chi^2$\\
\texttt{TVectorD Chi2List = Vfit->GetVtxChi2List();}// $\chi^2$ contribution of each track\\
\texttt{TVectorD NewPar = Vfit->GetNewPar(i);}// Updated i$^{th}$track parameters\\
\texttt{TMatrixDSym NewCov = Vfit->GetNewCov(i):}// Updated i$^{th}$ track covariance\\

We note that the instantiation of \texttt{VertexFit} performs only some basic initializations, while the actual fitting is triggered only when any of these methods is invoked: \texttt{GetVtx(), GetVtxCov(), GetVtxChi2() or GetVtxChi2List()}. 

The input tracks can also be added or removed incrementally as can be useful in pattern recognition combinatorics with the following methods:\\

\noindent
\texttt{Vfit->AddTrk(par, cov):} // Adds one charged track to the fit track list\\
\texttt{Vfit->AddTrk(par, cov, kFALSE):} // Adds one neutral track to the fit track list\\
\texttt{Vfit->RemoveTrk(i);} // Removes the i$^{th}$ track from the fit track list\\\

Finally the \texttt{VertexFit} class supports also including an external vertex constraint, for instance  an independent
 knowledge of the primary vertex position and width. Given the mean position, \texttt{Xpvc}, and the covariance matrix, \texttt{CovXpvc}, of this external constraint, one can include it in the fit with the call:\\

 \noindent
\texttt{Vfit->AddVtxConstraint(Xpvc, CovXpvc);}\\

N.B. when an external constraint is in place, vertex fits are allowed also with a single track in input.\\

Additional information after fitting can be obtained with the class \texttt{VertexMore}. Its constructor has in input the pointer to the vertex fit, \texttt{VertexFit* Vfit} and an optional \texttt{Bool\_t} parameter that requests the use of millimiters if \texttt{kTRUE}, as shown in the following:\\

\noindent
\texttt{VertexMore* VM = new VertexMore(Vfit);}\\
or\\
\noindent
\texttt{Bool\_t opt = kTRUE;\\
VertexMore* VM = new VertexMore(Vfit, opt);}\\

In the following the most relevant methods are described:\\

\noindent
\texttt{TVector3 GetMomentum(Int\_t i);}\tab// Momentum of the i$^{th}$track in the vertex\\
\texttt{TVector3 GetMomentumC(Int\_t i);}\tab// Covariance of the above\\  
\texttt{TVector3 GetTotalP();}\tab// Total momentum of vertex\\  
\texttt{TMatrixDSym GetTotalPcov();}\tab// Covariance of the above\\
\texttt{TVectorD GetXv();}\tab// Vertex position\\
\texttt{TMatrixDSym GetXvCov();}\tab// Covariance of the above\\
\texttt{TVectorD GetVpar();}\tab// Vertex track parameters\\
\texttt{TMatrixDSym GetVcov();}\tab// Covariance of the above\\
Mass constraints on a subset of the tracks in the vertex fit are initialized by calls like:\\
\noindent
\texttt{VM->AddMassConstraint(Double\_t Mass, Int\_t Ntr, Double\_t* masses, Int\_t* list);}\\
where \texttt{Mass} is the constraining mass value, \texttt{Ntr} is the number of tracks involved, \texttt{masses} is an array with the masses of the tracks involved and \texttt{list} is an array with the list of track numbers involved. This function can be called more than once if there is more than one disjoined set of tracks to constrain. The fit is then activated with\\

\noindent
\texttt{VM->MassConstrFit();}\\
that also updates all momenta and vertex positions and their covariance matrices, that can be accessed with the methods described above.

\section{Examples}\label{sec4}
Some examples on the use of the code are provided with the standard \texttt{DELPHES} distribution available in \texttt{GitHub}.
The most instructive ones are discussed in the following sections. It should be noted that all these examples assume input files generted with \texttt{Pythia8} and simulated with \texttt{DELPHES}.  It is important to ensure consistency with the generated files when beam constraints are used in the examples.

\subsection{Primary vertex determination}\label{subsec4.1}
The code for this example is located in \texttt{examples/ExamplePVtxFind.C} and is executed from \texttt{ROOT} by :\\
\noindent
\texttt{root>.X examples/ExamplePVtxFind.C("InputFile.root", 1000);}, where the second parameter is the number of events to process. A description of the algorithm and how the vertexing code is used is presented in the following.

The input is the full set of \texttt{NtrG} tracks in the event, that can be mapped to the track parameters, \texttt{pr[i]} and their covariance matrix, \texttt{cv[i]}. It is also assumed that there is an independent knowledge of the primary vertex with mean value \texttt{xpvc} and covariance \texttt{covpvc}, that wil be used as an external constraint. 

The algorithm first loops over all tracks and selects those sufficiently close to the known primary vertex:\\
\noindent
\texttt{// Skim tracks}\\
\texttt{Int\_t nSkim = 0;}\\
\texttt{Int\_t* nSkimmed = new Int\_t[NtrG];}\\
\texttt{TVectorD** PrSk = new TVectorD * [1];}\\
\texttt{TMatrixDSym** CvSk = new TMatrixDSym * [1];}\\
\texttt{Double\_t MaxChi2 = 9.;}\\
\texttt{for (Int\_t n = 0; n < NtrG; n++) \{}\\
\hspace*{20pt} \texttt{PrSk[0] = new TVectorD(*pr[n]);}\\
\hspace*{20pt} \texttt{CvSk[0] = new TMatrixDSym(*cv[n]);}\\
\hspace*{20pt} \texttt{// Vertex fit one track at a time}\\
\hspace*{20pt} \texttt{VertexFit* Vskim = new VertexFit(1,PrSk, CvSk);}\\
\hspace*{20pt} \texttt{// with external constraint}\\
\hspace*{20pt} \texttt{Vskim->AddVtxConstraint(xpvc, covpvc);}\\
\hspace*{20pt} \texttt{Double\_t Chi2One = Vskim->GetVtxChi2();}\\
\hspace*{20pt} \texttt{// Select depending on Chi2}\\
\hspace*{20pt} \texttt{if (Chi2One < MaxChi2) \{}\\
\hspace*{30pt} \texttt{nSkimmed[nSkim] = n;}\\
\hspace*{30pt} \texttt{nSkim++;\}}\\
\texttt{\}}\\

Then setup the primary vertex candidate fit with all selected tracks.\\

\noindent
\texttt{// Load all skimmed tracks}\\
\texttt{TVectorD** PrFit = new TVectorD * [nSkim];}\\
\texttt{TMatrixDSym** CvFit = new TMatrixDSym * [nSkim];}\\
\texttt{for (Int\_t n = 0; n < nSkim; n++) \{}\\
\hspace*{20pt} \texttt{PrFit[n] = new TVectorD(*pr[nSkimmed[n]]);}\\
\hspace*{20pt} \texttt{CvFit[n] = new TMatrixDSym(*cv[nSkimmed[n]]);\}}\\
\texttt{// Setup vertex fit}\\
\texttt{VertexFit* Vtx = new VertexFit(nSkim, PrFit, CvFit);}\\
\texttt{// add Constraint}\\
\texttt{Vtx->AddVtxConstraint(xpvc, covpvc);}\\

and remove iteratively the tracks that contribute most to the $\chi^2$ if above a given threshold.\\

\noindent
\texttt{//}\\
\texttt{// Remove tracks with large chi2}\\
\texttt{Double\_t MaxChi2Fit = 8.0;}\\
\texttt{Int\_t Nfound = nSkim;}\\
\texttt{const Int\_t MaxFound = 100; Double\_t Chi2LL[MaxFound];}\\
\texttt{Bool\_t Done = kFALSE;}\\
\texttt{while (!Done) \{}\\
\hspace*{10pt} \texttt{// Find largest Chi2 contribution}\\
\hspace*{10pt} \texttt{TVectorD Chi2List = Vtx->GetVtxChi2List();	// Contributions to Chi2}\\
\hspace*{10pt} \texttt{Chi2L = Chi2List.GetMatrixArray();}\\
\hspace*{10pt} \texttt{Int\_t iMax = TMath::LocMax(Nfound, Chi2L);}\\
\hspace*{10pt} \texttt{Double\_t Chi2Mx = Chi2L[iMax];	// Largest Chi2 contribution}\\
\hspace*{10pt} \texttt{if (Chi2Mx > MaxChi2Fit \&\& Nfound > 1) \{}\\
\hspace*{30pt} \texttt{// Remove bad track}\\
\hspace*{30pt} \texttt{Vtx->RemoveTrk(iMax);}\\
\hspace*{30pt} \texttt{Nfound--;\}}\\
\hspace*{10pt} \texttt{else \{Done = kTRUE;\}}\\
\texttt{\}}\\

After this selection \texttt{Vtx} is the final primary vertex.

\subsection{$B_s\rightarrow D_s\pi$}\label{subsec4.2}
The code for this example is located in \texttt{examples/VtxBs2DsPi.C} and is executed from \texttt{ROOT} by :\\
\noindent
\texttt{root>.L goBs.C+}\\
\texttt{root>goBs();}\\
\texttt{root>.X examples/VtxBs2DsPi.C("InputFile.root", 1000);}, where the second parameter is the number of events to process. The goal of this example is to first fit the decay $D_s^\pm\rightarrow  K^+K^-\pi^\pm$ and then use the $D_s$ track to fit the decay vertex $B_s\rightarrow D_s^-\pi^+$ or charge coniugate. A mass constraint is applied to the $D_s$ vertex fit to improve resolution.  A description of the algorithm and how the vertexing code is used is shown.\\

\noindent
\texttt{// Find Ds Vertex with mass constraint}\\
\texttt{// Load Ds tracks}\\
\texttt{TVectorD* tDsPar[nDsT];}\\
\texttt{TMatrixDSym* tDsCov[nDsT];}\\
\texttt{for(Int\_t k=0; k<nDsT; k++)\{}\\
\hspace*{10pt} \texttt{TVectorD par(5); TMatrixDSym cov(5);}\\
\hspace*{10pt} \texttt{TrkToVector(tDs[k], par, cov);}\\
\hspace*{10pt} \texttt{tDsPar[k] = new TVectorD(par);}\\
\hspace*{10pt} \texttt{tDsCov[k] = new TMatrixDSym(cov);}\\
\texttt{\}}\\
\texttt{// Fit Ds vertex}\\
\texttt{VertexFit* vDs = new VertexFit(nDsT, tDsPar, tDsCov);}\\
\texttt{Double\_t DsChi2 = vDs->GetVtxChi2();		// Ds fit Chi2}\\
\texttt{// More fitting}\\
\texttt{Bool\_t Units = kTRUE;		// Set to mm	}\\	
\texttt{VertexMore* VMDs = new VertexMore(vDs,Units);}\\
\texttt{// Mass constraint if requested}\\
\texttt{Bool\_t MCst = kTRUE;			// Mass constraint flag}\\
\texttt{if(MCst)\{}\\
\hspace*{10pt} \texttt{Double\_t DsMass = pBsDs->Mass;	// Ds mass}\\
\hspace*{10pt} \texttt{Double\_t DsMasses[nDsT]; Int\_t DsList[nDsT];}\\
\hspace*{10pt} \texttt{for(Int\_t k=0; k<nDsT; k++)\{}\\
\hspace*{30pt} \texttt{DsMasses[k] = pDs[k]->Mass;}\\
\hspace*{30pt} \texttt{DsList[k]   = k;}\\
\hspace*{10pt} \texttt{\}}\\
\hspace*{10pt} \texttt{VMDs->AddMassConstraint(DsMass,  nDsT,  DsMasses,  DsList);}\\
\hspace*{10pt} \texttt{VMDs->MassConstrFit();}\\
\texttt{\}}\\
\texttt{TVectorD rDv = VMDs->GetXv();			// Ds vertex}\\

Then use the $D_s$ track and the remaining $\pi$ to fit the $B_s$ vertex.\\

\noindent
\texttt{// Find Bs vertex}\\
\texttt{// Load Bs tracks}\\
\texttt{TVectorD* tBsPar[nBsT];}\\
\texttt{TMatrixDSym* tBsCov[nBsT];}\\
\texttt{TVectorD par(5); TMatrixDSym cov(5);}\\
\texttt{TrkToVector(tBs[0], par, cov);}\\
\texttt{tBsPar[0] = new TVectorD(par);			// Bs pion}\\	
\texttt{tBsCov[0] = new TMatrixDSym(cov);}\\
\texttt{tBsPar[1] = new TVectorD(VMDs->GetVpar());	// Ds from previous fit}\\
\texttt{tBsCov[1] = new TMatrixDSym(VMDs->GetVcov());}\\
\texttt{//}\\
\texttt{// Fit Bs vertex}\\
\texttt{VertexFit* vBs = new VertexFit(nBsT, tBsPar, tBsCov); // Bs vertex}\\
\texttt{Double\_t BsChi2 = vBs->GetVtxChi2();}\\

One could also use \texttt{VertexMore} for the $B_s$ vertex to extract more information or even apply an additional mass constraint on the $B_s$ if useful.

\subsection{$B_0\rightarrow K^0_S K^0_S$}\label{subsec4.3}
The code for this example is located in \texttt{examples/VtxB02KsKs.C} and is executed from \texttt{ROOT} by :\\
\noindent
\texttt{root>.L goKsKs.C+}\\
\texttt{root>goKsKs();}\\
\texttt{root>.X examples/VtxB02KsKs.C("InputFile.root", 1000);}, where the second parameter is the number of events to process. The goal of this example is to first fit the decays $K^0_s\rightarrow  \pi^+\pi^-$ and then use the $K^0_s$ tracks to fit the decay vertex $B_0\rightarrow K^0_s \,K^0_s$ or charge coniugate. A mass constraint is applied to the $K^0_s$ vertices fits to improve resolution.  A description of the algorithm and how the vertexing code is used is shown.\\

\noindent
\texttt{// Find first Ks vertex}\\	
\texttt{// Load 1st Ks tracks}\\	
\texttt{TVectorD* tKs1Par[nKsT];}\\	
\texttt{TMatrixDSym* tKs1Cov[nKsT];}\\	
\texttt{for(Int\_t k=0; k<nKsT; k++)\{}\\	
\hspace*{10pt} \texttt{TVectorD par(5); TMatrixDSym cov(5);}\\	
\hspace*{10pt} \texttt{TrkToVector(tKs1[k], par, cov);}\\	
\hspace*{10pt} \texttt{tKs1Par[k] = new TVectorD(par);}\\	
\hspace*{10pt} \texttt{tKs1Cov[k] = new TMatrixDSym(cov);}\\	
\texttt{\}}\\	
\texttt{// Fit 1st Ks vertex}\\	
\texttt{VertexFit* vKs1 = new VertexFit(nKsT, tKs1Par, tKs1Cov);}\\	
\texttt{Double\_t Ks1Chi2 = vKs1->GetVtxChi2();		// Ks fit Chi2}\\	
\texttt{// More fitting}\\	
\texttt{Bool\_t Units = kTRUE;		// Set to mm		}\\	
\texttt{VertexMore* VMKs1 = new VertexMore(vKs1,Units);}\\	
\texttt{// Mass constraint if requested}\\	
\texttt{Bool\_t MCst = kTRUE;			// Mass constraint flag}\\	
\texttt{if(MCst)\{}\\	
\hspace*{10pt} \texttt{Double\_t KsMass = pB0[0]->Mass;	// Ks mass}\\	
\hspace*{10pt} \texttt{Double\_t KsMasses[nKsT]; Int\_t KsList[nKsT];}\\	
\hspace*{10pt} \texttt{for(Int\_t k=0; k<nKsT; k++)\{}\\	
\hspace*{30pt} \texttt{KsMasses[k] = pKs1[k]->Mass;}\\	
\hspace*{30pt} \texttt{KsList[k]   = k;}\\	
\hspace*{10pt} \texttt{\}}\\	
\hspace*{10pt} \texttt{VMKs1->AddMassConstraint(KsMass,  nKsT,  KsMasses,  KsList);}\\	
\hspace*{10pt} \texttt{VMKs1->MassConstrFit();}\\	
\texttt{\}}\\	

A similar code is used to fit the second $K^0_s$ in the event. The $B_0$ vertex is then fit using the two neutral tracks:\\

\noindent
\texttt{// Find B0 vertex}\\
\texttt{// Load B0 tracks}\\
\texttt{TVectorD* tB0Par[nB0T];}\\
\texttt{TMatrixDSym* tB0Cov[nB0T];}\\
\texttt{tB0Par[0] = new TVectorD(VMKs1->GetVpar());	// 1st Ks previous fit	}\\
\texttt{tB0Cov[0] = new TMatrixDSym(VMKs1->GetVcov());}\\
\texttt{tB0Par[1] = new TVectorD(VMKs2->GetVpar());	// 2nd Ks previous fit}\\
\texttt{tB0Cov[1] = new TMatrixDSym(VMKs2->GetVcov());}\\
\texttt{//}\\
\texttt{// Fit B0 vertex}\\
\texttt{Bool\_t Charged[nB0T] = {kFALSE, kFALSE}; // Tag neutral tracks}\\
\texttt{VertexFit* vB0 = new VertexFit(nB0T, tB0Par, tB0Cov, Charged);}\\
\texttt{Double\_t B0Chi2 = vB0->GetVtxChi2();}\\
\texttt{// More fit information}\\
\texttt{VertexMore* VMB0 = new VertexMore(vB0,Units);}\\
\texttt{TVectorD rvmB0 = VMB0->GetXv();}\\

As in the previous case one could further improve the resolution by mass constraining the $B_0$ if useful.

\newpage
\begin{appendices}
\section{Steered parameter fit derivation}\label{secA}
Differentiating the $\chi^2$ shown in eq.~\ref{Chi2Steer} with respect to the track parameters and using the expansion of the track positions at the points $(s'_i, \vec{\alpha}'_i)$  we obtain:
\[ \frac{1}{2}\frac{\partial\chi^2}{\partial\,\vec{\alpha}_i}=C_i^{-1}(\vec{\alpha}_i-\vec{\alpha}^{\,0}_i)+A_i^t\vec{\lambda}_i  =
C_i^{-1}(\delta\vec{\alpha}_i-\delta\vec{\alpha}^{\,0}_i)+A_i^t\vec{\lambda}_i 
= 0 \] 
where $\delta\vec{\alpha}_i = \vec{\alpha}_i- \vec{\alpha}'_i$ and 
$\delta\vec{\alpha}^{\,0}_i = \vec{\alpha}^{\,0}_i- \vec{\alpha}'_i$.
The following derived relations will be used later:
\[ \vec{\alpha}_i = \vec{\alpha}^{\,0}_i-C_iA_i^t\vec{\lambda}_i  \]
and after mltiplication by  $A_i\,C_i$:
\[ A_i\,\delta\vec{\alpha}_i = A_i\,\delta\vec{\alpha}^{\,0}_i-(A_iC_iA_i^t)\vec{\lambda} 
= A_i\,\delta\vec{\alpha}^{\,0}_i-W_i^{-1}\vec{\lambda} \]
Differentiating with respect to the Lagrange multipliers returns the constraints:
\[ 0 = \vec{x}_i-\vec{x}_V = 
\vec{x}_i^{\,0}-\vec{x}_V +A_i\delta\vec{\alpha}_i+\vec{a}_i\delta\,s_i
=\vec{x}_i^{\,0}+A_i\,\delta\vec{\alpha}^{\,0}_i-\vec{x}_V-W_i^{-1}\vec{\lambda}_i+\vec{a}_i\delta\,s_i \] 
so:
\[ \vec{\lambda}_i = W_i\{\vec{x}_i^{\,0}+A_i\,\delta\vec{\alpha}^{\,0}_i-\vec{x}_V+\vec{a}_i\delta s_i\} \]
Differentiating by $s_i$ we can eliminate the dependence on $\delta s_i$ in the above:
\[ \frac{1}{2}\frac{\partial\chi^2}{\partial\,s_i}  = \vec{a}_i^t\vec{\lambda}_i = 0 \] so replacing $\vec{\lambda}_i$ from the above:
\[ \vec{a}_i^t\vec{\lambda}_i =
\vec{a}_i^tW_i\{\vec{x}_i^{\,0}+A_i\,\delta\vec{\alpha}^{\,0}_i-\vec{x}_V+\vec{a}_i\delta\,s_i\}=0 \]
leading to:
\[ \delta s_i = \vec{a}_i^tW_i\frac{\vec{x}_V-\vec{x}_i^{\,0}-A_i\,\delta\vec{\alpha}^{\,0}_i}{a_i} \]
where we have set $a_i = \vec{a}_i^tW_i\vec{a}_i$. Finally:
\begin{eqnarray}
\nonumber
\vec{\lambda}_i &=& W_i\{\vec{x}_i^{\,0}+A_i\,\delta\vec{\alpha}^{\,0}_i-\vec{x}_V
+\frac{\vec{a}_i\vec{a}_i^t}{a_i}W_i(\vec{x}_V-\vec{x}_i^{\,0}-A_i\,\delta\vec{\alpha}^{\,0}_i)\}\\
\nonumber
&=&\left(W_i-W_i \frac{\vec{a}_i\vec{a}_i^t}{a_i}W_i \right)(\vec{x}_i^{\,0}+A_i\,\delta\vec{\alpha}^{\,0}_i-\vec{x}_V) \\ \label{lambda}
&=& D_i(\vec{x}_i^{\,0}+A_i\,\delta\vec{\alpha}^{\,0}_i-\vec{x}_V)
\end{eqnarray}
We now differentiate by $\vec{x}_V$ obtaining:
\[ \frac{1}{2}\frac{\partial\chi^2}{\partial\,\vec{x}_V}  =
-\sum_{i=1}^N\vec{\lambda}_i = \sum_{i=1}^N  D_i(\vec{x}_V-\vec{x}_i^{\,0}-A_i\,\delta\vec{\alpha}^{\,0}_i)= 0 \]
that can be solved for $\vec{x}_V$ giving the solution:
\begin{equation}
\vec{x}_V = \left(\sum_{i=1}^N  D_i\right)^{-1}
\left(\sum_{i=1}^N  D_i(\vec{x}_i^{\,0}+A_i\,\delta\vec{\alpha}^{\,0}_i)\right) = D^{-1}\left(\sum_{i=1}^N  D_i(\vec{x}_i^{\,0}+A_i\,\delta\vec{\alpha}^{\,0}_i)\right)
\end{equation}
The corresponding error matrix on $\vec{x}_V$ is obtained by taking the average of the fluctuations of $\vec{x}_V$ due to those of the track parameters, $\vec{\alpha}^{\,0}_i$:
\begin{equation}
\begin{array}{lcl}
C_V = Cov(\vec{x}_V) &=& 
D^{-1}
\left(\sum_{ij}  D_iA_i
<\delta\vec{\alpha}_i^{\,0}\delta\vec{\alpha}_j^{\,0t}>A_i^{\,t}
D_j\right)
D^{-1}\\
&=&
D^{-1}
\left(\sum_{i}  D_i W_i^{-1} D_i\right)
D^{-1}
\end{array}
\end{equation}

The fitting procedure starts setting $\vec{\alpha}' = \vec{\alpha}^{\,0}$ and $s'$ to some smart guess of its value. After each iteration $\vec{\alpha}'$ is set to the latest value of $\vec{\alpha}$ obtained, given by:
\[ \vec{\alpha}_i = \vec{\alpha}_i^{\,0}-C_iA_i^{\,t}\vec{\lambda}_i \]
and $s'_i$ to the latest value of $s_i$ obtained given by:
\[  s_i = s'_i + \vec{a}_i^tW_i\frac{\vec{x}_V-\vec{x}_i^{\,0}-A_i\,\delta\vec{\alpha}^{\,0}_i}{a_i} \]

\section{Charged track formulas}\label{secB}
\subsection{Trajectory formulas}
In the point of minimum approach to the $z$-axis the 3D position and momentum are given by:
\begin{equation}
\left\{
\begin{array}{lcl}
x&=&x_0+[sin(s+\varphi_0)-sin(\varphi_0)]/(2C)\\
y&=&y_0-[cos(s+\varphi_0)-cos(\varphi_0)]/(2C)\\
z&=&z_0+\lambda\,s/(2C)
\end{array}\right.
\end{equation}
and 
\begin{equation}
\left\{
\begin{array}{lcl}
p_x&=&p_t\,cos(s+\varphi_0)\\
p_y&=&p_t\,sin(s+\varphi_0)\\
p_z&=&p_t\lambda
\end{array}\right.
\end{equation}
where
\begin{equation}
\begin{array}{ll}
\left\{
	\begin{array}{lcl}
	x_0&=&-D\,sin\,\varphi_0\\
	y_0&=& D\,cos\,\varphi_0\\
	z_0&=&z_0
	\end{array} \right. 
	&
\left\{
	\begin{array}{lcl}
	a&=&-0.2998BQ\;(T/m/GeV)\\
	\rho = 2C &=& a/p_t
	\end{array} \right.
\end{array}
\end{equation}
and $s$ is the angular displacement from the point of minimum approach to the $z$ axis (pma), $D$ is the signed transverse impact parameter, $\varphi_0$ the track direction at the pma, $z_0$ the $z$ coordinate at the pma, and $\lambda = p_z/p_t$ or the cotangent of the polar angle. The phase, $s$, at point $\vec{x}$ on the trajectory is given by:
\begin{equation}
s = sin^{-1}\{2C(x\,cos\,\varphi_0+y\,sin\,\varphi_0)\}
\end{equation}

In the following we show the derivatives of the track trajectory, $\vec{x}(\vec{\alpha},s)$, with respect to the track parameters, $\vec{\alpha}=(D,\,\varphi_0,\,C,\,z_0,\,\lambda)$, and the phase, $s$. They are shown in the following:

\begin{equation}
\begin{array}{ll}
  \vec{a} = \frac{\partial \vec{x}}{\partial s} =  
  \frac{1}{2C}\left(
  \begin{array}{c}
  cos(s+\varphi_0)\\sin(s+\varphi_0)\\\lambda
  \end{array}\right);\;
  &
  a = \vec{a}^{\,t}\,\vec{a} = \frac{1+\lambda^2}{4C^2}
\end{array}
\end{equation} 

\begin{equation}
A = \frac{\partial\vec{x}}{\partial\vec{\alpha}} = 
\left(
\begin{array}{ccccc}
-sin\,\varphi_0&-D\,cos\,\varphi_0+\frac{[cos(s+\varphi_0)-cos\,\varphi_0]}{2C}&-\frac{[sin(s+\varphi_0)-sin\,\varphi_0]}{2C^2}&0&0\\
cos\,\varphi_0&-D\,sin\,\varphi_0+\frac{[sin(s+\varphi_0)-sin\,\varphi_0]}{2C}&\frac{[cos(s+\varphi_0)-cos\,\varphi_0]}{2C^2}&0&0\\
0&0&-\lambda\,s/(2C^2)&1&s/(2C)
\end{array} \right)
\end{equation}

Similarly for the track momentum:
\begin{equation}
\frac{\partial\vec{p}}{\partial\vec{\alpha}} = \left(
\begin{array}{rrrrr}
0&-p_y&-p_x/C&0&0\\
0& p_x&-p_y/C&0&0\\
0&0&-p_z/C&0&p_t\\
\end{array}\right)
\end{equation}

and
\begin{equation}
\frac{\partial\vec{p}}{\partial s} = (-p_y, \,p_x,\,0)
\end{equation}

\subsection{Track parameters from $\vec{x},\,\vec{p}$}
We provide the basic formulas to obtain the track parameters in cylindrical geometry given a point on the track, $\vec{x}$,  and the momentum, $\vec{p}$, in that point. 
Two of the parameters, $\lambda$ and $C$, depend only on the momentum:
\begin{equation}
\lambda = p_z/p_\perp;\;\;C = a/(2p_\perp)
\end{equation}
The transverse impact parameter, $D$, is given by:
\begin{equation}
D = \frac{1}{a}(T-p_\perp);\;
T = \sqrt{p^2_\perp-2a(xp_y-yp_x)+a^2(x^2+y^2)}
\end{equation}
The angle $\varphi_0$ is given by:
\begin{equation}
cos\,\varphi_0 = \frac{p_x+ay}{T};\;sin\,\varphi_0 = \frac{p_y-ax}{T}
\end{equation}
or
\begin{equation}
tan\,\varphi_0 = \frac{p_y-ax}{p_x+ay}
\end{equation}
Finally $z_0$:
\begin{equation}
z_0 = z-\lambda s/2C
\end{equation}

The derivatives of the track parameters relative to both $\vec{x}$ and $\vec{p}$ are reported in the following.
\begin{equation}
\frac{\partial\lambda}{\partial\vec{x}} = (0,\,0,\,0);\;\frac{\partial\lambda}{\partial\vec{p}} =
\left(-\frac{p_xp_z}{p_\perp^3},\,-\frac{p_yp_z}{p_\perp^3}, \frac{1}{p_\perp}\right)
\end{equation}
\begin{equation}
\frac{\partial C}{\partial\vec{x}} = (0,\,0,\,0);\;\frac{\partial C}{\partial\vec{p}} =
\frac{a}{2}\left(-\frac{p_x}{p_\perp^3},\,-\frac{p_y}{p_\perp^3},\,0\right)
\end{equation}
\begin{equation}
\frac{\partial T}{\partial\vec{x}} = 
\left(-\frac{a}{T}(p_y-ax),\,\frac{a}{T}(p_x+ay),\,0\right);\;
\frac{\partial T}{\partial\vec{p}} = \left(\frac{p_x+ay}{T},\,\frac{p_y-ax}{T},\,0\right)
\end{equation}
\begin{equation}
\frac{\partial D}{\partial\vec{x}} = \frac{1}{a}\frac{\partial T}{\partial\vec{x}};\;
\frac{\partial D}{\partial\vec{p}} = \left(\frac{1}{a}\left(\frac{\partial T}{\partial p_x}-
\frac{p_x}{p_\perp}\right),\, \frac{1}{a}\left(\frac{\partial T}{\partial p_y}-
\frac{p_y}{p_\perp}\right),\,0\right)
\end{equation}
\begin{equation}
\frac{\partial\varphi_0}{\partial\vec{x}} = -\frac{a\,cos^2\,\varphi_0}{p_x+ay}\,
\left(1,\,tan\,\varphi_0,\,0\right)
\end{equation}
\begin{equation}
\frac{\partial\varphi_0}{\partial\vec{p}} = \frac{cos^2\,\varphi_0}{p_x+ay}\,
\left(-tan\,\varphi_0,\,1,\,0\right)
\end{equation}
The derivative of $z_0$ is calculated after noting that an alternate expression for the phase, $s$, is given by:
\begin{equation}
s = tg^{-1}\frac{p_y}{p_x}-\varphi_0
\end{equation}
then:
\begin{equation}
\frac{\partial z_0}{\partial\vec{x}} = 
\left(\frac{\lambda}{2C}\frac{\partial\varphi_0}{\partial x},\,\frac{\lambda}{2C}\frac{\partial\varphi_0}{\partial y},\,1\right) =
\left(\frac{p_z}{a}\frac{\partial\varphi_0}{\partial x},\,\frac{p_z}{a}\frac{\partial\varphi_0}{\partial y},\,1\right) 
\end{equation}
\begin{equation}
\frac{\partial z_0}{\partial\vec{p}} = 
\left(
\frac{p_z}{a}\left(\frac{p_y}{p_t^2}+\frac{\partial \varphi_0}{\partial p_x}\right),\,
\frac{p_z}{a}\left(-\frac{p_x}{p_t^2}+\frac{\partial \varphi_0}{\partial p_y}\right),\,
-s/a
\right)
\end{equation}

\section{Neutral track formulas}\label{secC}
The track trajectory equation for  neutral tracks is a straight line. For consistency with the previous parameterization we find the trajectory equation by setting $C = 0$ in the charged track equations obtaining:
\begin{equation}
\begin{array}{ll}
\left\{
\begin{array}{lcl}
x&=&x_0+s\,cos\varphi_0\\
y&=&y_0+s\,sin\varphi_0\\
z&=&z_0+s\,\lambda
\end{array}\right.\mbox{\;\;where\;\;}
&
\left\{
\begin{array}{lcl}
x_0&=&-D\,sin\varphi_0\\
y_0&=&D\,cos\varphi_0\\
s&=&\sqrt{R^2-D^2}
\end{array}\right.
\end{array}
\end{equation}
we note that in this case the parameter $s$ is not angle, but the distance from the pma, and is given by:
\begin{equation}
s = x\,cos\,\varphi_0+y\,sin\,\varphi_0
\end{equation}
Alternatively, as a function of the radius, $R$, the neutral track equation can be writtten as:
\begin{equation}
\left\{
\begin{array}{lcl}
\varphi(R)&=&\varphi_0+sin^{-1}(D/R)\\
z(R)&=&z_0+\lambda\,\sqrt{R^2-D^2}
\end{array}\right.
\end{equation}

There are only 4 parameters to define the trajectory, however we include also a 5$^{th}$ parameter, $p_t$, to keep track of the neutral track momentum. The final set of parameters describing a neutral is threrefore $\vec{\alpha}\,=\, (D,\,\varphi_0,\,p_t,\,z_0,\,\lambda)$. The derivatives of the trajectory with respect to the track parameters is:
\begin{equation}
\frac{\partial\vec{x}}{\partial\vec{\alpha}} = \left(
\begin{array}{ccccc}
-sin\varphi_0&-D\,cos\varphi_0-s\,sin\varphi_0&0&0&0\\
cos\varphi_0&-D\,sin\varphi_0+s\,cos\varphi_0&0&0&0\\
0&0&0&1&s
\end{array}\right) = \left(
\begin{array}{ccccc}
-sin\varphi_0&-y&0&0&0\\
cos\varphi_0&x&0&0&0\\
0&0&0&1&s
\end{array}\right)
\end{equation}

Derivatives of the trajectory with respect to $s$ are given by:
\begin{equation}
\frac{\partial\vec{x}}{\partial\,s}=
(cos\varphi_0,\,sin\varphi_0,\,\lambda)
\end{equation}
Neutral track momenta are constant and given by:
\begin{equation}
\vec{p} = (p_t\,cos\,\varphi_0,\,p_t\,sin\,\varphi_0,\,p_t\lambda) 
\end{equation}
with trivial derivatives with respect to track parameters:
\begin{equation}
\frac{\partial \vec{p}}{\partial\vec{\alpha}} = \left(
\begin{array}{rrrrr}
0&-p_y&cos\varphi_0&0&0\\
0&p_x&sin\varphi_0&0&0\\
0&0&\lambda&0&p_t
\end{array}\right)
\end{equation}

\subsection{Neutral track parameters from $\vec{x}$ and $\vec{p}$}
The track direction is easily obtained from the momentum:
\begin{equation}
\left\{
\begin{array}{clc}
cos\varphi_0&=&p_x/p_t\\
sin\varphi_0&=&p_y/p_t\\
\lambda&=&p_z/p_t
\end{array}\right. \mbox{\;\;where\;\;}
p_t=\sqrt{p_x^2+p_y^2}
\end{equation}
The parameters $D$ and $z_0$ can be obtained by eliminating $s$ in the trajectory equation:
\begin{equation}
D=y\,cos\varphi_0-x\,sin\varphi_0,\;s=y\,sin\varphi_0+x\,cos\varphi_0,\;z_0=z-\lambda\,s
\end{equation}
Derivatives of the phase, $s$, with respect to $\vec{x}$:
\begin{equation}
\frac{\partial s}{\partial \vec{x}}=
(cos\varphi_0, \,sin\varphi_0, \,0)
\end{equation}
and with respect to the momentum, $\vec{p}$:
\begin{equation}
\frac{\partial s}{\partial\vec{p}}=\left(
\frac{-D\,sin\varphi_0}{p_t},\,\frac{D\,cos\varphi_0}{p_t},\,0 \right)
\end{equation}
Derivatives of the  track parameters with respect to the starting point $\vec{x}$ and the momentum $\vec{p}$ are given by:
\begin{equation}
\frac{\partial\vec{\alpha}}{\partial\vec{x}}=\left(
\begin{array}{ccccc}
-sin\varphi_0&0&0&-\lambda\,cos\varphi_0&0\\
cos\varphi_0&0&0&-\lambda\,sin\varphi_0&0\\
0&0&0&1&0
\end{array}\right)
\end{equation}
and
\begin{equation}
\frac{\partial\vec{\alpha}}{\partial\vec{p}}=\left(
\begin{array}{ccccc}
\frac{s\,sin\varphi_0}{p_t}&-\frac{sin\varphi_0}{p_t}&cos\varphi_0&\frac{\lambda}{p_t}(s\cdot cos\varphi_0+D\,sin\varphi_0)&-\frac{\lambda}{p_t}\,cos\varphi_0\\
\frac{-s\,cos\varphi_0}{p_t}&\frac{cos\varphi_0}{p_t}&sin\varphi_0&
\frac{\lambda}{p_t}(s\cdot sin\varphi_0-D\,cos\varphi_0)&-\frac{\lambda}{p_t}\,sin\varphi_0\\
0&0&0&-s/p_t&1/p_t
\end{array}\right)
\end{equation}

\end{appendices}
\newpage

\end{document}